\documentclass[12pt,preprint]{aastex}
\usepackage{apjfonts, emulateapj5, natbib}
\input psfig.tex

\def\aa{{A\&A}}

\def\aj{{AJ}}
\def\al{$\alpha$}
\def\bet{$\beta$}

\def\apj{{ApJ}}
\def\apjs{{ApJS}}
\def\asec{$^{\prime\prime}$}

\def\etal{{et al. }}

\def\hal{H$\alpha$}
\def\ha{H$\alpha$}
\def\hb{H$\beta$}

\def\kms{km s$^{-1}$}
\def\lamb{$\lambda$}
\def\lax{{$\mathrel{\hbox{\rlap{\hbox{\lower4pt\hbox{$\sim$}}}\hbox{$<$}}}$}}
\def\gax{{$\mathrel{\hbox{\rlap{\hbox{\lower4pt\hbox{$\sim$}}}\hbox{$>$}}}$}}
\def\simlt{\lower.5ex\hbox{$\; \buildrel < \over \sim \;$}}
\def\simgt{\lower.5ex\hbox{$\; \buildrel > \over \sim \;$}}
\def\lum{erg s$^{-1}$}
\def\mbh{{$M_{\rm BH}$}}
\def\micron{{$\mu$m}}
\def\mnras{{MNRAS}}
\def\nat{{Nature}}
\def\pasp{{PASP}}

\def\percm2{cm$^{-2}$}

\def\solmass{$M_\odot$}

\def\heii{\ion{He}{2}}

\def\oiii{[\ion{O}{3}]}

\def\feii{\ion{Fe}{2}}
\def\feiii{\ion{Fe}{3}}
\def\mgii{\ion{Mg}{2}}
\def\civ{\ion{C}{4}}
\def\ciii{\ion{C}{3]}}

\def\lbol{$L_{{\rm bol}}$}
\def\ledd{$L_{{\rm Edd}}$}

\def\mbh{{$M_{\rm BH}$}}
\newcommand{\chisq}{\ensuremath{\chi^2}}

\slugcomment{To appear in {\it The Astrophysical Journal}.}
\shorttitle{Line Profiles of Quasars}
\shortauthors{HO ET AL.}

\begin{document}

\title{Simultaneous Ultraviolet and Optical Emission-line Profiles of 
Quasars: Implications for Black Hole Mass Determination\altaffilmark{1}}

\author{Luis C. Ho\altaffilmark{2}, Paolo Goldoni\altaffilmark{3,4}, Xiao-Bo 
Dong\altaffilmark{2,5}, Jenny E. Greene\altaffilmark{6}, and Gabriele 
Ponti\altaffilmark{3,7}}

\altaffiltext{1}{Based on observations collected at the European Organisation for Astronomical Research in the Southern Hemisphere, Chile, under program 086.B-0320(A).}

\altaffiltext{2}{The Observatories of the Carnegie Institution for Science, 
813 Santa Barbara Street, Pasadena, CA 91101, USA}

\altaffiltext{3}{Laboratoire Astroparticule et Cosmologie, 10 rue A. Domon 
et L. Duquet, 75205 Paris Cedex 13, France}

\altaffiltext{4}{DSM/IRFU/Service d'Astrophysique, CEA/Saclay, 91191 
Gif-sur-Yvette, France}

\altaffiltext{5}{Key laboratory for Research in Galaxies and Cosmology,
The University of Sciences and Technology of China, Chinese Academy of 
Sciences, Hefei, Anhui 230026, China}

\altaffiltext{6}{Department of Astrophysical Sciences, Princeton University, 
Peyton Hall, Ivy Lane, Princeton, NJ 08544, USA}

\altaffiltext{7}{School of Physics and Astronomy, University of Southampton, 
Highfield, SO17 1BJ, UK}

\begin{abstract}
The X-shooter instrument on the VLT was used to obtain spectra of seven 
moderate-redshift quasars simultaneously covering the spectral range 
$\sim$3000 \AA\ to 2.5 \micron.  At $z \approx 1.5$, most of the prominent broad emission 
lines in the ultraviolet to optical region are captured in their rest frame.  
We use this unique dataset, which mitigates complications from source 
variability, to intercompare the line profiles of \civ\ \lamb1549, \ciii\ 
\lamb1909, \mgii\ \lamb2800, and \ha\ and evaluate their implications for 
black hole mass estimation.  We confirm that \mgii\ and the Balmer lines share 
similar kinematics and that they deliver mutually consistent black hole mass 
estimates with minimal internal scatter (\lax 0.1 dex) using the latest virial 
mass estimators.  Although no virial mass formalism has yet been calibrated 
for \ciii, this line does not appear promising for such an application because 
of the large spread of its velocity width compared to lines of both higher and 
lower ionization; part of the discrepancy may be due to the difficulty of 
deblending \ciii\ from its neighboring lines.  The situation for \civ\ is 
complex and, because of the 
limited statistics of our small sample, inconclusive.  On the one hand, 
slightly more than half of our sample (4/7) have \civ\ line widths that 
correlate reasonably well with \ha\ line widths, and their respective black 
hole mass estimates agree to within $\sim$0.15 dex.  The rest, on the other 
hand, exhibit exceptionally broad \civ\ profiles that overestimate virial 
masses by factors of 2--5 compared to \ha.  As \civ\ is widely used to study 
black hole demographics at high redshifts, we urgently need to revisit our 
analysis with a larger sample.
\end{abstract}


\section{Motivation}

Ever since their discovery in the early 1960s, quasars have attracted attention
not only as laboratories for exploring extreme regimes in astrophysics, but 
also because they serve as useful beacons for probing the interstellar and 
intergalactic medium.  With the growing appreciation that central black holes 
(BHs) are ubiquitous and inextricably tied to galaxy formation and evolution 
(Cattaneo et~al. 2009, and references therein), quasars have gained even 
greater prominence in their unique role as markers of vigorous BH growth out 
to the highest accessible redshifts (Mortlock et~al. 2011; Treister et~al. 
2011).  

A key development comes from our ability to estimate BH masses for broad-lined 
(type~1) active galactic nuclei (AGNs), including quasars, using simple 
parameters that can be extracted from their rest-frame ultraviolet (UV) and 
optical spectra.  The basic premise is that the BH mass can be approximated by 
the virial product $M_{\rm BH} = f R_{\rm BLR} \Delta V^2/G$, where 
$R_{\rm BLR}$, the radius of the broad-line region (BLR), can be estimated from 
the radius-luminosity relation calibrated through reverberating mapping 
experiments (Kaspi et~al. 2000, 2005; Bentz et~al. 2009), $\Delta V$ 
is the velocity dispersion of the line-emitting gas that can be measured from 
the widths of the broad emission lines, $G$ is the gravitational constant, and 
$f$ is a geometric factor of order unity that accounts for the poorly 
constrained geometry and kinematics of the BLR.  A variety of approaches have 
been employed to calibrate the virial method for deriving $M_{\rm BH}$ using 
single-epoch spectra, using different emission lines for different 
redshift regimes.  H\bet\ is normally the line of choice for low-redshift 
objects (Kaspi et~al. 2000), as it is the principal line for which most 
reverberation mapping experiments have been done to date, although under some 
circumstances H\al\ is preferable to H\bet\ (Greene \& Ho 2005b).  At 
intermediate redshifts, 0.75 \lax\ $z$ \lax\ 2, McLure \& Jarvis (2002) 
introduced a formalism based on \mgii\ \lamb2800, while \civ\ \lamb1549 is 
the only viable option for quasars at $z$ \gax\ 2 (Vestergaard 2002).  With 
the increasing availability of near-infrared (NIR) spectroscopy, these 
restrictions can now be circumvented by observing the rest-frame optical lines 
out to high redshift, thereby minimizing the additional uncertainties incurred 
through the extra layers of intermediary cross-calibrations.  Nevertheless, 
access to NIR spectroscopy is still far from routine, and all extant, large 
spectral databases used for statistical analyses of AGNs continue to rely on 
optical surveys.  

A number of studies have investigated the robustness of different broad 
emission lines commonly used to estimate BH masses.  The general consensus is 
that \mgii\ serves as a reasonably effective substitute for \hb, the local 
``standard.'' H\bet\ and \mgii, both low-ionization lines, are thought to 
arise from gas with common physical conditions and presumably from similar 
locations with similar kinematics.  This expectation appears 

\begin{figure*}[t]
\centerline{\psfig{file=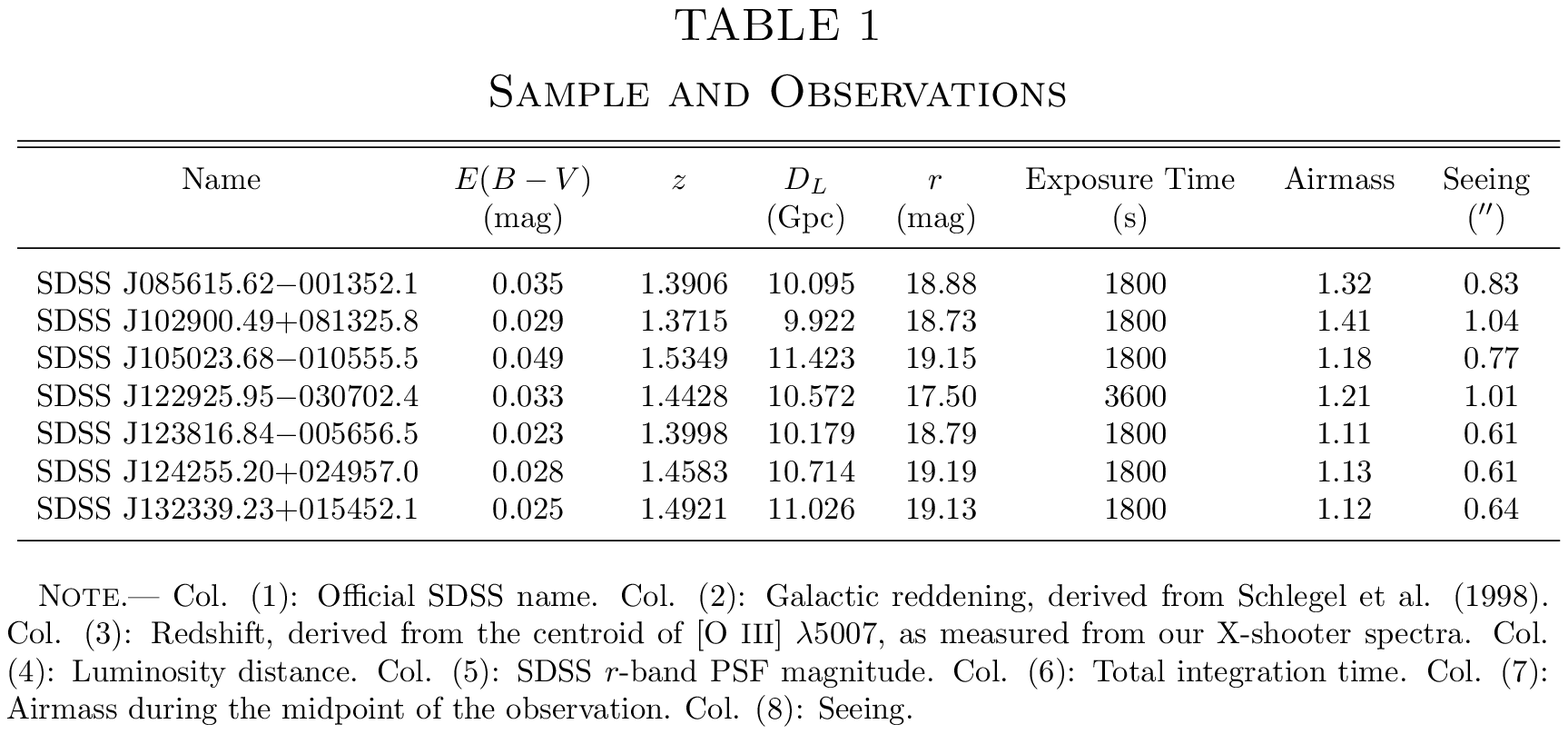,width=16.5cm,angle=0}}
\end{figure*}

\noindent
to hold to a 
first approximation (McLure \& Dunlop 2004; Salviander et~al. 2007; McGill et 
al.  2008; Shen et~al. 2008), and whatever small residual differences there 
might be appear correctable with empirical prescriptions (Onken \& Kollmeier 
2008) or more refined spectral analysis (Rafiee \& Hall 2011).  \civ, on the 
other hand, turns out to be more problematic.  While some contend that \civ\ 
can be calibrated to deliver useful mass estimates (Vestergaard 2002; Warner 
et~al. 2003; Vestergaard \& Peterson 2006; Kelly \& Bechtold 2007; Dietrich 
et~al. 2009; Assef et~al. 2011), others sound a more pessimistic note.  Baskin 
\& Laor (2005) systematically compared the profiles of \civ\ and \hb\ for 
low-redshift ($z$ \lax\ 0.5) quasars for which they could locate published 
optical and archival space-based UV spectra.  The two lines agree poorly.  Not 
only does the width of \civ\ show large and apparently systematic deviations 
from \hb, but, as long known (Gaskell 1982; Tytler \& Fan 1991), the profile 
of \civ\ is often highly blueshifted and asymmetric, casting serious doubt as 
to whether the line properly traces gravitationally bound gas\footnote{
Vestergaard \& Peterson (2006) reassessed Baskin \& Laor's analysis and 
concluded that the mismatch between \civ\ and \hb, though certainly 
substantial, is not as severe has had been claimed.}.  Netzer et~al. (2007), 
Sulentic et~al. (2007), and Shen \& Liu (2012) arrive at a similar conclusion 
from analysis of moderate-redshift ($z \approx 2$) quasars for which they 
secured rest-frame Balmer line measurements using near-IR spectroscopy; the 
widths of the Balmer lines exhibit little, if any, correlation with the widths 
of \civ.

AGNs vary.  Larger variations usually occur at shorter wavelengths and in 
lines of higher ionization.  An important limitation of most of the previous 
studies comes from the fact that the rest-frame UV and optical lines were 
observed non-simultaneously, often separated widely apart in time (timescales 
of months to years), making any comparison between them inherently uncertain.  
The width of \civ\ in quasars, for instance, changes up to $\sim$30\% on 
timescales of weeks to months (Wilhite et~al. 2006a).  Fortunately, 
variability has only a relatively minor impact on BH masses derived from 
single-epoch spectra.  According to Wilhite et~al. (2006b), Denney et~al. 
(2009), and Park et al. (2012) variability affects the mass estimates only at 
the level of $\sim$0.1 dex, which is small, but significant, compared to the 
$0.3-0.4$ dex error budget that still plagues the various mass estimators 
(Vestergaard \& Peterson 2006; McGill et~al. 2008).  Still, to achieve a 
better understanding of how the different lines most commonly used for BH mass 
estimation compare to each other, it would be highly desirable to perform a 
comparative analysis using a set of {\it simultaneous}\ rest-frame 
UV--optical spectra of quasars.  This is the main goal of this paper.

\vskip 0.3cm
\psfig{file=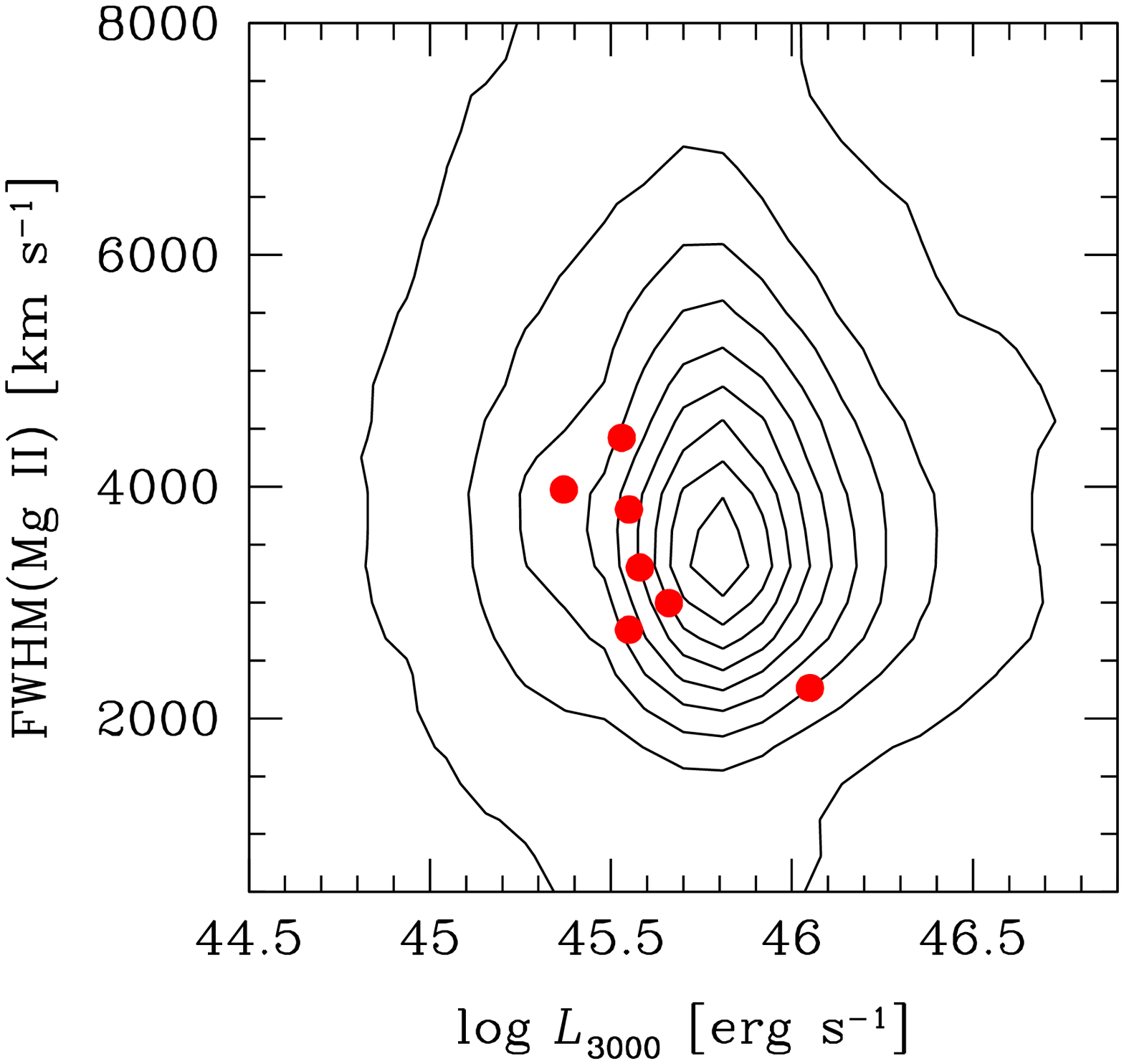,width=8.75cm,angle=0}
\figcaption[fig1.ps]{
Distribution of the continuum luminosity at 3000 \AA\ and the FWHM of the 
broad component of \mgii\ \lamb2800 for 11,015 SDSS DR7 quasars with 
$z = 1.4-1.6$ (Shen et al. 2011). The objects observed with X-shooter are 
plotted as red points.  The values of FWHM from SDSS were reduced by 0.05 dex 
to account for the systematic difference in the method used to fit \mgii\ 
(Shen et al. 2011).
\label{fig1}}

\vskip 1.5cm

\begin{figure*}[t]
\centerline{\psfig{file=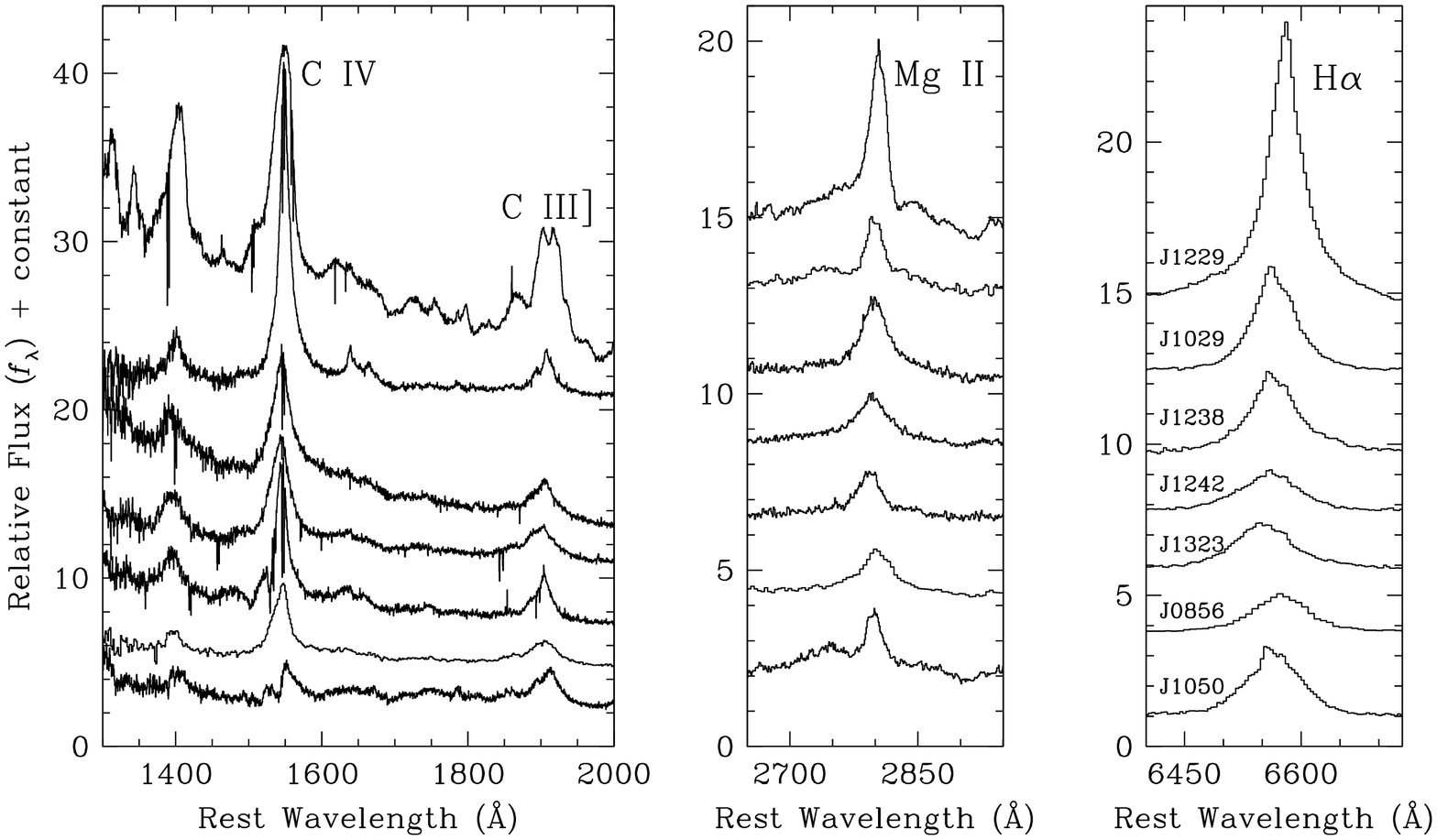,width=17.5cm,angle=270}}
\figcaption[fig2.ps]{
X-shooter spectra of our sample, shifted to the rest-frame of the objects.
We only show the spectral regions that contain the strongest emission lines
used for the profile analysis.  The individual objects have been shifted by
arbitrary constant additive offsets to improve the clarify of the plot.
\label{fig2}}
\end{figure*}

Distance-dependent quantities are calculated assuming the following 
cosmological parameters: $H_0 = 100~h = 71$~\kms~Mpc$^{-1}$, $\Omega_{m} = 
0.26$, and $\Omega_{\Lambda} = 0.74$ (Komatsu \etal\ 2009).

\section{Observations and Data Reduction}

We selected seven relatively bright ($r \approx 18-19$ mag) quasars from the 
Seventh Data Release of the Sloan Digital Sky Survey (SDSS DR7; Abazajian 
et~al.  2009).  As we are interested in simultaneous coverage of \civ\ 
\lamb1549 to \ha, we chose the targets to have $z \approx 1.5$.  During the 
selection process, we inspected the existing SDSS spectra to ensure that the 
objects have unambiguous broad emission lines suitable for estimating BH 
masses.  We avoided sources with obvious broad absorption features, but aside 
from this, we did not apply any other selection criteria.  Figure~1 compares 
our sample with the distribution of \mgii\ line widths (FWHM) and 3000 \AA\ 
continuum luminosity for 11,015 $z = 1.4-1.6$ quasars contained in the DR7 
catalog of Shen et al. (2011).  Our objects are on average slightly less 
luminous than the peak of the DR7 distribution, but there are no obvious 
biases.  The one apparent outlier with relatively narrow lines is 
SDSS~J122925.95$-$030702.4, which was intentionally included for comparison 
because it has characteristics similar to narrow-line Seyfert 1 (NLS1) 
galaxies.  Thus, even though our sample is small and somewhat ill-defined, it 
nonetheless represents an unbiased subset of luminous ($L_{3000} > 10^{45.4}$ 
\lum), moderate-redshift, optically selected quasars.

The observations were made using X-shooter (Vernet et~al. 2011), a three-arm, 
single-object echelle spectrograph that started operations in 2009 October on 
the VLT.  The instrument covers simultaneously the wavelength range from 
$\sim$3000 \AA\ to 2.40 \micron, in three arms: UVB ($\Delta \lambda = 
3000 - 5500$ \AA), VIS ($\Delta \lambda = 5500$ \AA\ -- 1.02 \micron), and NIR 
($\Delta \lambda =$ 1.02--2.40 \micron).  The data were taken within the 
framework of the French Guaranteed Time under program 086.B-0320(A) (PI 
G. Ponti) and took place on 2010 February 18 UT.  A summary of the 
observations is given in Table~1.  

For our observations we used slit widths of 1\farcs3, 1\farcs2, and 1\farcs2,
respectively, for the three arms, resulting in resolving powers of $R \equiv
\lambda/\Delta \lambda \approx$ 4000, 6700, and 4300.  The slits were aligned 
along the parallactic angle.  The observations consist of four separate 
exposures of 450~s each, for a total of 1800~s, for all sources except 
SDSS~J122925.95$-$030702.4, which was observed for 900~s per exposure for a 
total of 3600~s.  The exposures were taken while nodding the object along the 
slit with an offset of 5\asec\ between exposures in a standard ABBA sequence.  
Every observation was preceded by an observation of an A0~V telluric standard 
star at similar airmass. The night was not 
photometric\footnote{\tt archive.eso.org/asm/ambient-server?site=paranal} due 
to the presence of thin clouds. A high, variable humidity was present with 
precipitable water vapor\footnote{\tt archive.eso.org/bin/qc1$\_$cgi?action=qc1$\_$plot$\_$table\&table=ambient$\_$PWV} varying between 5 and 7.5 mm,
rather high for Paranal but not exceptional for this period of the year in 
Chile (so-called ``Bolivian winter'').

We processed the spectra using version 1.1.0 of the X-shooter data reduction 
pipeline (Goldoni et~al. 2006), which performed the following actions. 
As usual in processing nodding observations, the raw frames
taken at different positions were differenced with each other to obtain 
two frames (i.e. A$-$B and B$-$A), on which cosmic ray hits were detected and 
corrected using the method developed by van~Dokkum (2001).  The frames were 
then divided by a master flat field obtained using day-time flat 

\begin{figure*}[t]
\centerline{\psfig{file=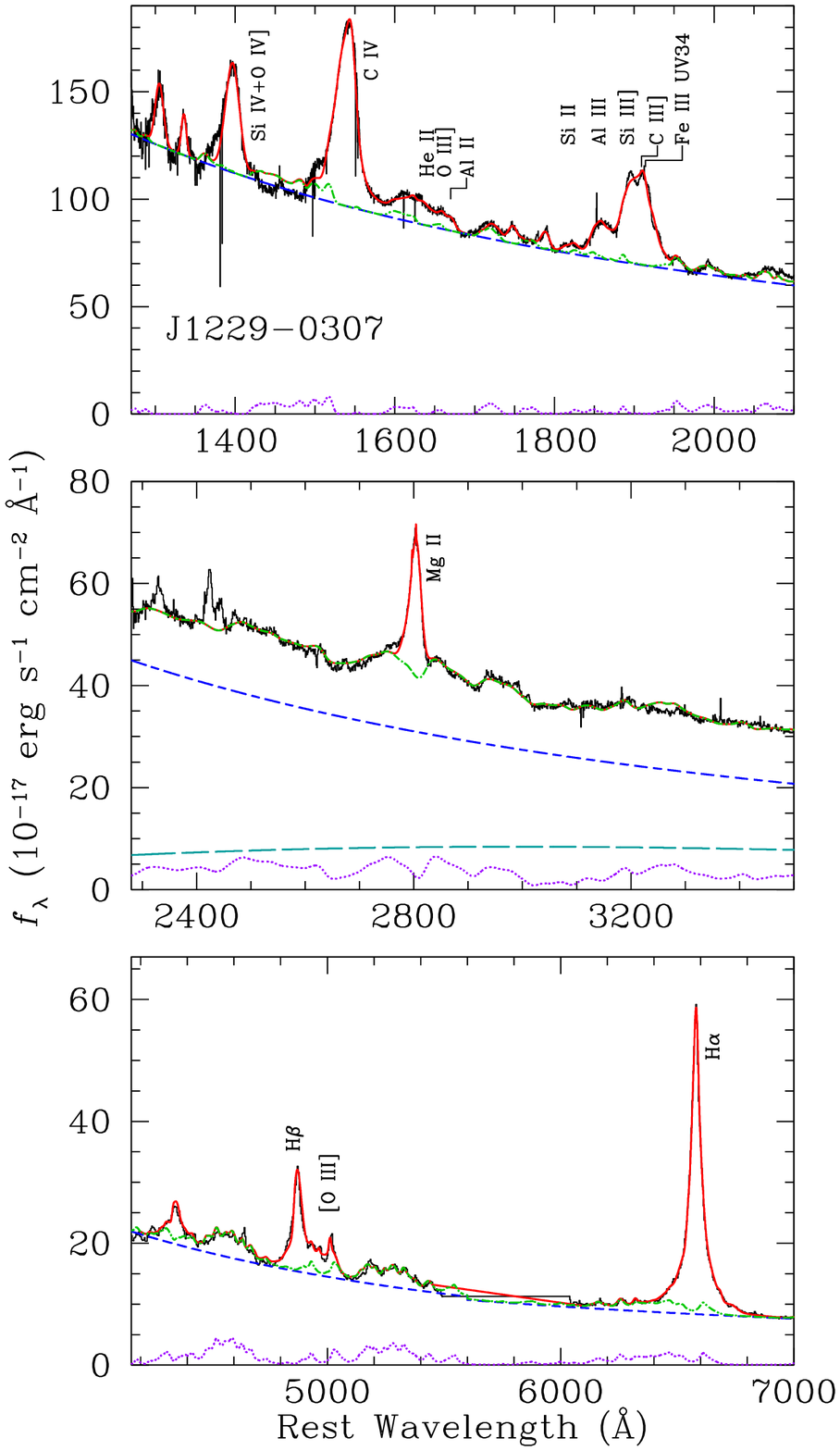,height=8.5in,angle=0}}
\figcaption[fig3.ps]{
Spectral fitting for one of the objects in our sample,
SDSS~J1229$-$0307.  The original data, corrected for Galactic reddening, are
plotted in black histograms.  The fitted components include a featureless
power law (dashed blue), the Balmer continuum (long dashed turquoise), and iron
emission (dotted purple), which altogether constitute the pseudocontinuum
(dot-dashed green). The final model, including fits to the principal emission
lines, is shown in solid red.
\label{fig3}}
\end{figure*}
\vskip 0.3cm
\vskip 1.3cm

\begin{figure*}[t]
\centerline{\psfig{file=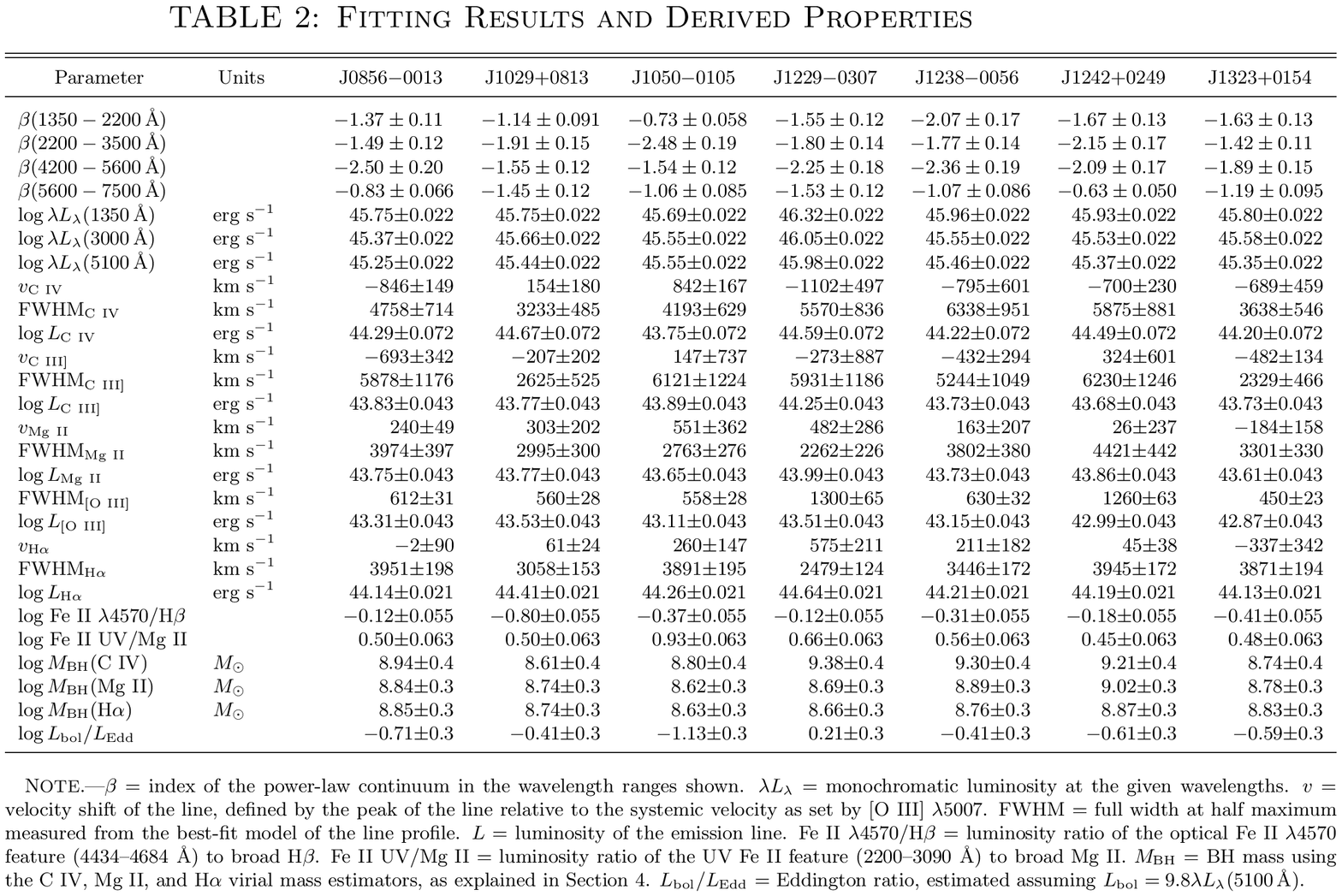,width=17.5cm,angle=-270}}
\end{figure*}

\noindent
field 
exposures with halogen lamps.  The orders were extracted and rectified in 
wavelength space using a wavelength solution previously obtained from 
calibration frames.  The resulting rectified orders were then shifted and 
added to superpose them, thus obtaining the final two-dimensional spectrum.  
The orders were then merged, and, in the overlapping regions, the merging was 
weighted by the errors that were propagated during the process. From the 
resulting two-dimensional, merged spectrum a one-dimensional spectrum was 
extracted at the source's position.  The one-dimensional spectrum with the 
corresponding error file and bad pixel map is the final product of the 
reduction.

To perform flux calibration, we used different procedures for the UVB data and 
for the VIS-NIR data.  In the UVB band we used an observation of the flux 
standard GD~71  (Bohlin 2007) taken in the beginning of the night. We reduced 
the data using the same steps as above, but in this case we subtracted the sky 
emission lines using the method of Kelson (2003). The extracted spectrum was 
divided by the flux table of the star from the CALSPEC HST database\footnote{
\tt www.stsci.edu/hst/observatory/cdbs/calspec.html} to produce the response 
function, which was then applied to the spectrum of the science targets.  For 
the VIS and NIR arms, we used A0V stars as both flux and telluric standards. 
We extracted the A0V spectra with the same procedure used for the flux 
standard and used these spectra to apply telluric corrections and flux 
calibrations simultaneously, using the package {\tt Spextool} (Vacca et~al. 
2003).  We verified that the final spectra of the three arms were compatible 
in the common wavelength regions and then adjusted the mean continuum flux in 
the overlap range between 4000 \AA\ and 9000 \AA\ to be consistent with the 
SDSS spectra, which have a more reliable absolute flux calibration.  Figure~2 
shows the calibrated, rest-frame spectra of the quasars observed in our 
program.  

\section{Spectral Fitting}

We correct the X-shooter and SDSS spectra for Galactic extinction using the 
extinction map of Schlegel et~al.\ (1998) and the reddening curve of 
Fitzpatrick (1999). The spectra are transformed into the rest frame using the 
redshift as determined from the peak of the best-fit model (see below) for the 
\oiii\,$\lambda5007$ line.  To improve on the absolute flux calibration, we 
rescale the X-shooter spectra to the flux density level of the SDSS spectra 
over the wavelength region where the two overlap.  Here we present a brief 
description of the spectral fitting, which is based on \chisq-minimization 
using the Levenberg--Marquardt technique within the MPFIT package 
(Markwardt 2009).

The analysis for the optical spectra closely follows the methodology for 
decomposition of AGN spectra described in Dong et~al. (2008).  We do not 
correct for starlight contamination, which is expected to be negligible 
for quasars in our luminosity range ($L_{3000} > 10^{45.4}$ \lum).  As the 
broad emission lines, particularly the \feii\ multiplets, are so broad and 
strong that they merge together and essentially leave no line-free wavelength 
regions, we fit simultaneously the nuclear 

\vskip 0.3cm
\begin{figure*}[t]
\centerline{\psfig{file=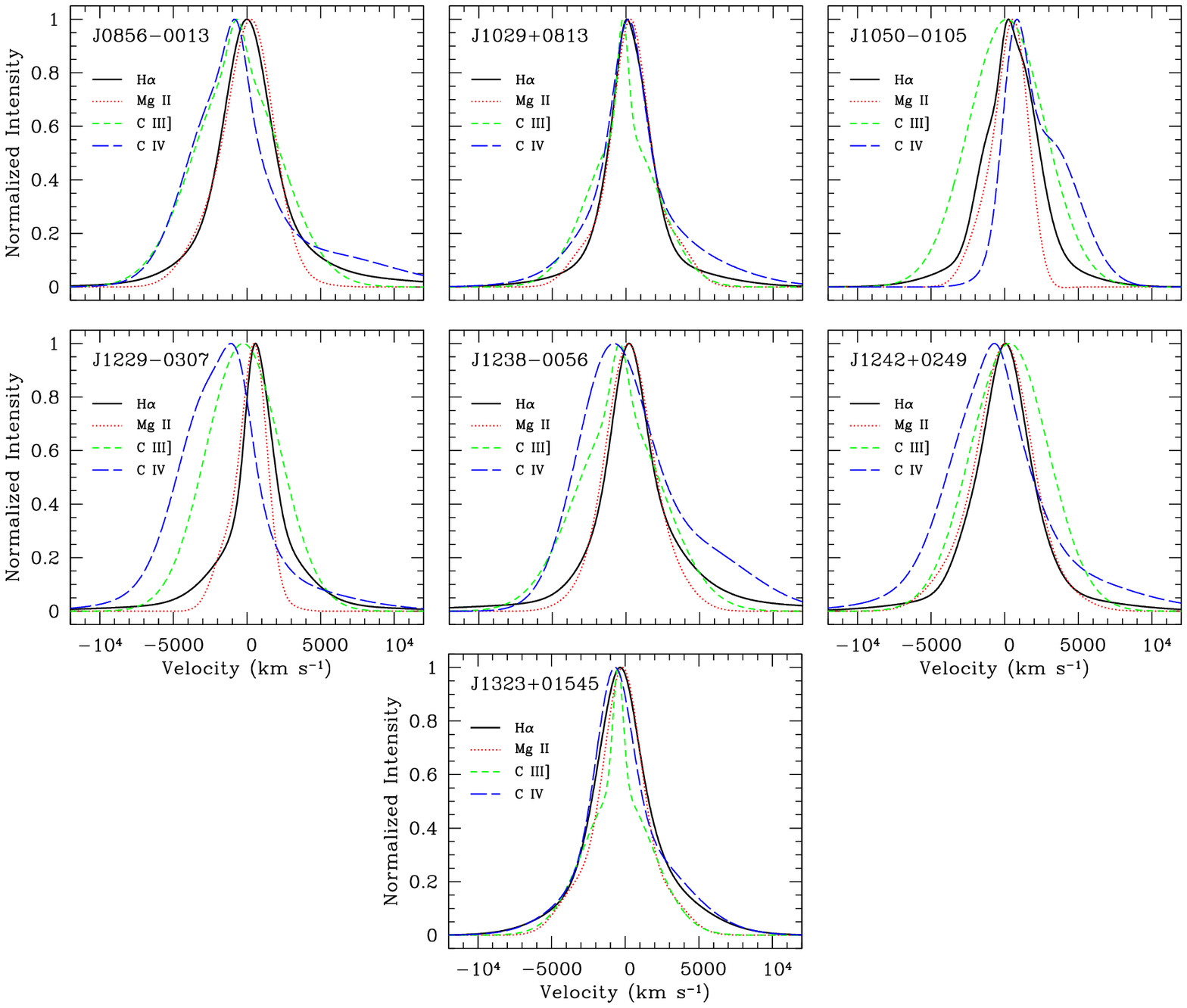,width=17.0cm,angle=0}}
\figcaption[fig4.ps]{
Model line profiles, shown on a velocity scale, for \ha\ \lamb6563 (black, 
solid), \mgii\ \lamb2800 (red, dotted), \ciii\ \lamb1909 (green, dashed), 
and \civ\ \lamb1549 (blue, long dashed).  The zero point of the velocity
scale is referenced with respect to \oiii\ \lamb5007.
\label{fig4}}
\end{figure*}

\noindent
continuum, the \feii\ multiplets,
and other emission lines.  The featureless continuum of type 1 AGNs is not
well described by a single power law when a large range of wavelengths
is considered (e.g., Vanden Berk et al. 2001).  We approximate it by a broken
power law, with free indices for the \ha\ and \hb\ regions.  The optical \feii\
emission is modeled with two separate sets of analytical spectral templates,
one for the broad-line system and the other for the narrow-line system,
constructed from measurements of I\,Zw\,1 by V\'eron-Cetty et~al. (2004).
Within each system, the respective set of \feii\ lines is assumed to have no
relative velocity shifts and the same relative strengths as those in I~Zw~1.
Emission lines are modeled as multiple Gaussians.  Following Dong et~al.
(2011), we assume that the broad \feii\ lines have the same profile as broad
\ha.  In our data set, the \hb\ region is significantly noisier than the \ha\ 
region. In our subsequent analysis we will use \ha\ in lieu of \hb; Greene \& 
Ho (2005b) have shown that the broad component of \ha\ is on average slightly 
narrower than \hb, but the two closely track each other.  All narrow emission 
lines are fitted with a single Gaussian, except for the \oiii\ $\lambda 
\lambda4959,5007$ doublet, each of which is modeled with two Gaussians, one 
accounting for the line core and the other for a possible blue wing seen in 
many objects (Greene \& Ho 2005a).

For the near-UV spectra, we focus our analysis on the region around
\mgii\,$\lambda \lambda$2796,\,2803, following the method described in Wang 
et~al. (2009).  Here the pseudocontinuum consists not only of a local power-law 
continuum and an \feii\ template, but also an additional component for the 
Balmer continuum.  The \feii\ emission is modeled with the semi-empirical 
template for I~Zw~1 generated by Tsuzuki et~al. (2006). To match the width 
and possible velocity shift of the \feii\ lines, we convolve the template with 
a Gaussian and shift it in velocity space. As in Dietrich et~al. (2002), the 
Balmer continuum is assumed to be produced in partially optically thick clouds 
with a uniform

\vskip 0.3cm
\begin{figure*}[t]
\centerline{\psfig{file=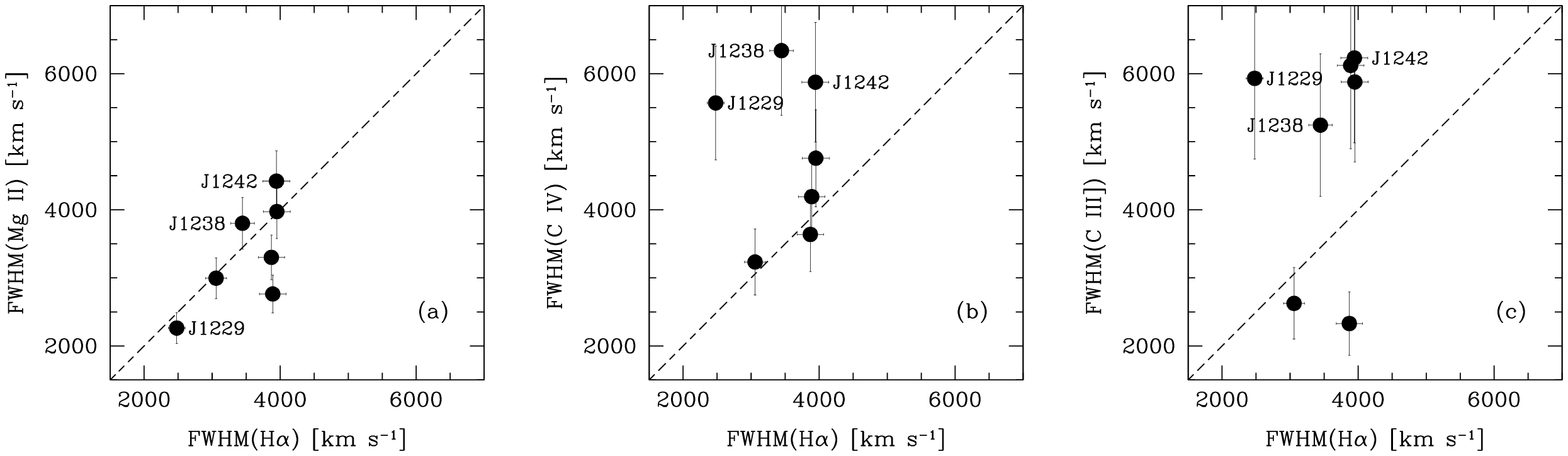,width=18.5cm,angle=270}}
\figcaption[fig5.ps]{
Comparison of FWHM of H\al\ with FWHM of (a) \mgii, (b) \civ, and (c) \ciii.
The error bars represent measurement uncertainties as given in
Table 2 (see Section 3).
\label{fig5}}
\end{figure*}
\vskip 0.3cm

\noindent
temperature.  Each of the lines of the \mgii\ doublet is modeled
with two components, one broad and the other narrow. The broad component is fit
with a truncated, five-parameter Gauss-Hermite series; a single Gaussian is
used for the narrow component.  The two doublet lines are assumed to have
identical profiles, a fixed separation set to the laboratory value, and
a flux ratio \lamb2796/\lamb2803 set to be between 2:1 and 1:1 (Laor et al.
1997).

The fitting for the UV region is performed with a modified version of a code
initially written and kindly provided by Jian-Guo~Wang. We employ the UV
\feii~$+$~\feiii\ template for I~Zw~1 generated by Vestergaard \& Wilkes
(2001).  In light of the moderate signal-to-noise ratio of our data, we adopt
the same scaling factor for the \feii\ and \feiii\ emission; this assumption
seems adequate from visual inspection of the fits.  As in Vestergaard \& 
Wilkes (2001), the power-law continuum and iron emission (both constituting 
the pseudocontinuum) are fit to the emission-line--free spectral regions 
from 1300 \AA\ to 2200 \AA.  After the pseudocontinuum is subtracted, we 
concentrate our fits on two regions: (1) $\sim1350-1700$~\AA, which contains 
the blend of \ion{Si}{4} and \ion{O}{4}] at $\sim$1400 \AA, 
\civ\,$\lambda1549$, \heii\,$\lambda1640$, O\,III]\,$\lambda1663$, and 
Al~II\,$\lambda1670$, and (2) $\sim 1800-2000$~\AA, which covers 
\ion{Si}{2}\,$\lambda1817$, \ion{Al}{3}\,$\lambda\lambda1855,\,1863$, 
\ion{Si}{3}]\,$\lambda1892$, \ciii\ $\lambda1909$, and the Fe\,III UV34 
triplet at $\sim$1914 \AA.  We model the emission lines with multiple Gaussian
components, using the minimum number necessary to achieve a satisfactory fit 
within the signal-to-noise constraints of the data.  The \civ\ profile is fit 
with three Gaussians, taking care to avoid narrow absorption features when 
present (as in J1238$-$0056 and J1323$+$0154; see Figure~2).  For the \ciii\ 
region, \ion{Si}{2}, \ion{Al}{3}, \ion{Si}{3}], and Fe\,III are each fit with 
a single Gaussian, while \ciii\ itself is fit with two Gaussians, which in 
most cases yield better residuals than a single Gaussian.  This suffices for 
our purposes.  Our primary objective is to obtain a robust characterization of 
the profile of \civ\ and \ciii, not to achieve a detailed model for every line 
in this tremendously complicated spectral region.

Figure~3 illustrates the fits for one of objects, and the results of the fits 
for the entire sample are summarized in Table~2.  The emission-line 
luminosities, velocity shifts (peak of the line; $v$), and velocity widths 
(FWHMs) are measured from the best-fit models of the line profiles, except for 
the case of \mgii, whose line peak and FWHM are measured from the model of the 
single doublet line, \mgii\ $\lambda$2796. The strength of the optical 
\feii\,$\lambda4570$ emission is integrated over the wavelength range 
4434--4684~\AA, and that for UV \feii\ is integrated over the range 2200--3090 
\AA. For all measured emission-line fluxes, we regard the values as reliable 
detections when they have greater than $3\,\sigma$ significance.  As the 
instrumental resolution (FWHM $\approx 45-75$ \kms) is small compared to the 
measured line widths (FWHM $\approx 500-6000$ \kms), no correction for 
instrumental resolution has been applied to the line widths.

We follow the bootstrap method of Dong et al. (2008; their Section 2.5) to 
estimate measurement uncertainties.  The typical 1~$\sigma$ errors on the 
fluxes of the strong lines considered in this paper are quite small, \lax 10\% 
(see Dong et al.  2008, 2011; Wang et al. 2009).  We adopt 10\% errors for the 
fluxes of \mgii, \ciii, and \oiii\ and 5\% for \ha.  The only exception is 
\civ, whose final flux depends on the exact procedure adopted to fit the ``red 
shelf'' (region $\sim 1600-1700$ \AA; see, e.g., Fine et~al. 2010).  We assume 
that the red shelf is intrinsic to \civ; excluding it reduces the inferred line 
flux by $\sim$10--25\%.  To be conservative, we adopt an error of 18\% for the 
flux of \civ.  The FWHM values are best measured for \ha\ (error 5\%), 
followed by \mgii\ (error 10\%), \civ\ (error 15\%), and \ciii\ (error 
15\%--25\%).  \ciii\ is particularly problematic because of severe blending 
with \ion{Al}{3}\,$\lambda\lambda1855,\,1863$, \ion{Si}{3}]\,$\lambda1892$, 
and Fe\,III UV34.  


\section{Results}

Figure~4 shows the final, fitted profiles for the sample, normalized by their 
amplitude and plotted on a velocity scale.  A direct comparison among the 
line widths is given in Figure~5.  Several trends are immediately obvious.  

\begin{enumerate}

\item
H\al\ shows, by far, the most symmetric profiles and the least amount of 
systemic velocity shift.  This is not unexpected, given the overall 
similarity between \ha\ and \hb\ profiles (Greene \& Ho 2005b), reinforcing 
the commonly adopted view that the Balmer lines offer the most reliable BH 
virial mass estimators (but see Vestergaard et al. 2011 for a different point 
of view).  

\item
Consistent with other studies (e.g., McLure \& Dunlop 2004; Shen et~al. 2008; 
Shen \& Liu 2012), we find that \mgii\ on average serves as a reasonably decent proxy for the Balmer lines, at least for five out of the seven objects in our 
sample.  The two exceptions are J1050$-$0105 and 

\vskip 0.3cm
\begin{figure*}[t]
\centerline{\psfig{file=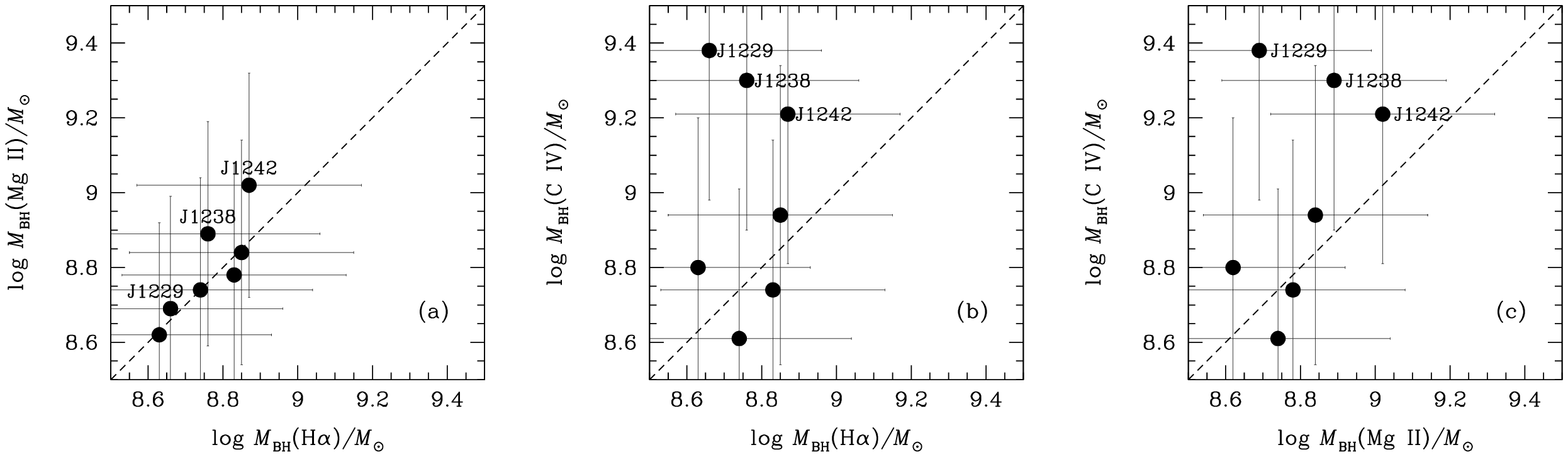,width=18.5cm,angle=270}}
\figcaption[fig6.ps]{
Comparison of $M_{\rm BH}$ derived using (a) \mgii\ and \hal, (b)
\civ\ and \hal, and (c) \civ\ and \mgii.  The error bars reflect the
statistical uncertainties for the BH mass estimators (see Section 4).
\label{fig6}}
\end{figure*}
\vskip 0.3cm

J1229$-$0307, whose \mgii\
lines are distinctly narrower than \ha\ and somewhat asymmetric.

\item
Among the five objects that show better match between \mgii\ and \ha,
the lines agree best near the core (FWHM) but often diverge noticeably in
the wings.  \ha\ generally has more extended wings than \mgii.  This implies
that comparison between virial mass estimators for \mgii\ and the Balmer lines
depends sensitively on the adopted measure of line width.  The agreement should
be reasonably close for the FWHM (Figure~5(a); mean and standard deviation
$\langle$FWHM(\mgii)/FWHM(\ha)$\rangle = 0.95\pm 0.14$ \kms), but the line
dispersion, more sensitive to the line wings, will be systematically higher
for the Balmer lines compared to \mgii.

\item
The behavior of \civ\ is much more perplexing.  Apart from the presence of a
prominent extended red wing---a model-dependent feature sensitive to the 
adopted fitting procedure (Fine et~al. 2010)---the line peak exhibits a range 
of velocity shifts, from $\sim$800 \kms\ to $-1100$ \kms.  A direct comparison 
between \civ\ and \ha\ line widths reveals two clusters of points 
(Figure~5(b)).  Four of the quasars have \civ\ line widths that are roughly 
consistent with those of \ha\ 
(mean and standard deviation $\langle$FWHM(\civ)/FWHM(\ha)$\rangle = 1.07\pm 
0.11$ \kms), whereas the other three (J1229$-$0307, J1238$-$0056, and 
J1242$+$0249) are significantly broader, 
$\langle$FWHM(\civ)/FWHM(\ha)$\rangle 
= 1.86\pm 0.38$ \kms.  Both are difficult to understand in the context of a 
radially stratified BLR with a velocity field dominated by Keplerian rotation,
wherein $\Delta V \propto R_{\rm BLR}^{1/2}$.  Not many AGNs have been 
reverberation mapped in \civ, but the handful for which adequate data exist 
show that \civ\ arises from a region that is a factor of $\sim 2$ more compact 
than \hb\ (e.g., Korista et~al. 1995; Onken \& Peterson 2002). If these 
results apply to higher luminosity quasars, they imply that FWHM(\civ) 
$\approx$ $\sqrt 2$FWHM(\ha) [for simplicity, we assume FWHM(\ha) = FWHM(\hb)].
Our sample, albeit small, does not seem to conform to this simple expectation.
The three outliers with unusually high FWHM(\civ)/FWHM(\ha) share one 
common characteristic: the peak of the \civ\ is significantly blueshifted, 
by $-$700 to $-$1100 \kms.  However, not all objects with \civ\ blueshifts
show enhanced FWHM(\civ)/FWHM(\ha); two of the other four objects also show
blueshifts of comparable magnitude.  Nor is excess \civ\ width uniquely 
associated with any obvious AGN property.  While J1229$-$0307, with its 
relatively narrow lines [FWHM(\ha) $\approx$ 2500 \kms], strong \feii\ 
emission (log \feii\ \lamb4570/\hb\ = $-0.12$), and high Eddington 
ratio\footnote{We assume bolometric luminosity \lbol\ = $9.8 
\lambda L_\lambda({\rm 5100\,\AA})$ (McLure \& Dunlop 2004), 
Eddington luminosity $L_{\rm Edd}\,=\,1.26 \times 10^{38} \left(M_{\rm BH}/
M_{\odot}\right)$ \lum, and BH mass $M_{\rm BH}$ estimated from \ha.} 
(log \lbol/\ledd\ = 0.21) qualify it as a NLS1, whose \civ\ line may be 
particularly problematic for mass determination (Vestergaard et al. 2011), the 
other two outliers do not.

\item
\ciii\ poses an even greater challenge to understand.  While there are no 
large velocity shifts, the width of \ciii\ appears to be completely erratic
(Figure~5(c)).  Two of the objects have unusually narrow cores that result in 
FWHM(\ciii)/FWHM(\ha) $<$ 1, and the rest are characterized by 
FWHM(\ciii)/FWHM(\ha) $\gg$ 1.  Counterintuitively, FWHM(\ciii) $>$ FWHM(\civ) 
in four out of the seven sources.  This agrees with the results of Brotherton 
et~al. (1994), Jiang et~al. (2007), and Greene et~al. (2010), but is 
inconsistent with the study of Shang et~al. (2007), who generally find 
FWHM(\ciii) \lax\  FWHM(\civ).  Using a much larger sample, Shen \& Liu (2012)
find no obvious offset between FWHM(\ciii) and FWHM(\civ), although the two 
are poorly correlated.  The discrepancy associated with \ciii\ may arise, 
at least in part, from the uncertainty in deblending the line from its 
surrounding contaminating features.

\end{enumerate}

How do these results impact BH mass determinations?  Figure~6 graphically 
illustrates the answer.  We estimate BH masses using the \civ-based formalism
of Vestergaard \& Peterson (2006), the \mgii-based formalism of Vestergaard \& 
Osmer (2009), and the \ha-based formalism of Greene \& Ho (2005b), as updated 
by Xiao et~al. (2011) to account for the latest BLR size-luminosity relation of 
Bentz et~al. (2009).  For consistency with our adopted \civ\ and \mgii\ mass 
estimators, we further adjust Xiao et~al.'s prescription so that its virial 
coefficient $f$ matches the value advocated by Onken et~al. (2004).  We do not 
consider \ciii\ because there is currently no mass estimator based on this 
line, and we feel discouraged by the line width comparison presented above.
The BH masses derived from these virial mass estimators typical have 
statistical uncertainties of $\sim 0.3-0.4$ dex.  For concreteness, we adopt
an error bar of 0.4 dex for \mbh(\civ) and 0.3 dex for both \mbh(\mgii) and 
\mbh(\ha).

Vestergaard et al. (2011) stress the importance of using mass estimators 
calibrated on the same mass scale for a proper comparison between masses 
derived from different lines.  Both the \civ\ and \mgii\ masses are tied to a 
common scale based on reverberation-mapped AGNs.  Vestergaard \& Osmer, 
however, model the broad \mgii\ line as a single component, whereas we follow 
Wang et al. (2009) and treat \mgii\ as a doublet, which results in slightly 
narrower line widths.  These two methods yield \mgii\ line widths that differ 
on average only by $\sim 0.05$ dex (Shen et al. 2011);  we scale up all of our 
\mgii\ line widths by this constant factor to account for the minor systematic 
offset.  To date the \ha\ BH mass estimator of Greene \& Ho (2005b) has not 
been calibrated directly against the reverberation-mapped AGNs. Fortunately, 
the analysis of Shen et al. (2011) indicates that the \ha-based masses show 
only a mean offset of 0.08 dex with respect to Vestergaard \& Peterson's 
\hb-based masses, which, like the \civ\ and \mgii\ masses, are calibrated to 
the same scale tied to the reverberation-mapped AGNs.  For the purposes of 
this paper, we will not worry about this small discrepancy, which does not 
impact any of our main conclusions.

As foreshadowed by the line width analysis, \ha\ and \mgii\ deliver reasonably 
consistent mass estimates, which for our sample spans a small range around 
$M_{\rm BH} \approx 10^{8.7\pm0.2}$ \solmass.  For our choice of mass 
estimators, \mbh(\mgii) agrees very well with \mbh(\ha): $\langle$log 
\mbh(\mgii)$-$log \mbh(\ha)$\rangle \approx 0.03\pm0.07$ dex.  By contrast, 
the comparison between \mbh(\civ) and \mbh(\ha) is less clear-cut.  The three 
objects (labeled in Figure~6) with anomalously broad, blueshifted \civ\ 
profiles all have \mbh(\civ) in excess of $10^{9}$ \solmass, deviating from 
\mbh(\hal) by 0.34 dex to as much as 0.72 dex.  The most discrepant object is 
the NLS1 J1229$-$0307, confirming Vestergaard et al.'s (2011) suspicion that 
\civ\ masses are especially unreliable for this class of AGNs, but the other 
two show no obvious warning signs as to why \civ\ should misbehave.  All three 
\civ\ outliers appear relatively normal in the \mgii\ versus \ha\ comparison 
(Figure~6(a)).  The remaining four objects fare better: 
$\langle$log \mbh(\civ)$-$log \mbh(\ha)$\rangle \approx 0.01\pm0.14$ dex.
While this may be regarded as reassuring confirmation that \civ-based masses 
can be trusted in at least \emph{some} objects, the difficulty is that, in the 
absence of independent evidence from lower ionization lines (\mgii, \hb, or 
\ha), we have no means of forecasting \emph{which} objects are reliable or 
not.  This result is unsettling, especially in light of the very large masses 
($M_{\rm BH} \approx 10^{9}-10^{10}$ \solmass) routinely inferred for 
high-redshift quasars and the astrophysical implications attached to them.

\section{Summary}

The widths of broad emission lines in active galaxies, when combined with 
physical dimensions inferred from the size-luminosity relation empirically 
calibrated from reverberation mapping experiments, provide a powerful and 
efficient means of estimating BH masses for large samples of sources detected 
at all redshifts.  Over the past decade, a number of virial BH mass estimators 
have been devised and extensively used to determine BH masses, facilitating 
a wide range of investigations on BH demographics and AGN physics.  Despite 
their popularity, however, there are nagging doubts as to the reliability 
of these mass estimators.  While calibrations based on low-ionization lines, 
especially \ha\ and \hb, are reasonably secure, higher redshift observations 
often depend on UV lines that are less well understood.  The most contentious 
mass estimator is that based on \civ\ \lamb 1549, the workhorse for studies of 
high-redshift quasars, as the kinematics of this line may be significantly 
affected by winds and other non-virial motions.  \mgii\ appears to be safer, 
but even it may not be completely immune to systematic biases.  

Spectral variability complicates the comparison between different mass 
estimators, if they derive from spectra taken at different times.  To mitigate 
this effect, we have undertaken an experiment to acquire spectra of a small 
sample of seven moderate-redshift quasars that \emph{simultaneously} cover the 
rest-frame UV through optical spectral regions (1300--7500 \AA).  This dataset 
is enabled by the unique capabilities of the X-shooter instrument on the VLT, 
which delivers simultaneous spectra from $\sim$3000 \AA\ to 2.5 \micron.  At 
$z \approx 1.5$, this allows us to access the principal broad emission lines 
from \civ\ to \ha.  

In accord with other studies, we find that \mgii\ and the Balmer lines (this 
study uses \ha\ instead of \hb) have similar velocity widths near the 
core (FWHM) of their profiles, but \ha\ generally has more extended, higher 
velocity wings than \mgii.  \mgii-based and \ha-based BH masses agree to 
better than $\sim 0.1$ dex.  The \ciii\ \lamb1909 line widths are difficult to 
interpret.  Contrary to naive expectations, FWHM(\ciii) $>$ FWHM(\civ) in most 
of our objects, but some also have unusually narrow lines, narrower than even 
those of \ha.   While these discrepancies can perhaps be attributed to the 
difficulties of line deblending in the highly crowded spectral region near 
$\sim$1900 \AA, it appears that it would be challenging to devise a robust 
BH virial mass estimator using \ciii.  The verdict on \civ\ is mixed.  While 
roughly half of the \civ-based masses are reasonably consistent with those 
derived from \ha, the others are systematically high by factors of 2--5.  These 
extreme outliers not only have unusually broad \civ\ line widths, but their 
line peaks are all systematically blueshifted by several hundred to a 
thousand \kms, suggesting that a significant fraction of the emission arises
from outflowing or dynamically unrelaxed gas.  However, systemic \civ\ 
blueshifts are a common feature in AGNs, and they do not appear to be a clean 
predictor of which objects show deviant \civ\ line widths or masses; two of 
the objects whose \civ-based masses agree with those derived from \ha\ also 
have systematic \civ\ blueshifts of comparable magnitude.  We are thus left in 
an uncomfortable predicament: in the absence of independent confirmation
from lower ionization lines, we do not know, a priori, which \civ\ profiles
provide more accurate mass estimates.

To end on a more positive note, we emphasize that our results clearly suffer 
from small-number statistics.  Other investigators (e.g., Vestergaard \& 
Peterson 2006; Greene et al. 2010; Assef et al. 2011) are more optimistic
that \civ\ can be used to study BH demographics at high redshifts.  It would
be important to secure simultaneous, or at least near-contemporaneous, 
rest-frame UV-optical spectra for a larger, better defined sample of
AGNs to revisit the issues raised in this study.  Because high-luminosity AGNs 
vary on time scales of weeks to months, complete simultaneity in spectral 
coverage (such as those presented here) is desirable, but not truly necessary.
Near-contemporaneous observations, taken within a span of a few days (e.g., 
during the same observing run or coordinated between two different 
telescopes), would suffice.

\acknowledgements
We thank an anonymous referee for helpful criticisms.  We are grateful to 
Jian-Guo~Wang for sharing his code for fitting the UV spectra.  This work was 
supported by the Carnegie Institution for Science.  GP acknowledges support 
via an EU Marie Curie Intra-European Fellowship under contract no. 
FP7-PEOPLE-2009-IEF-254279.


\end{document}